\documentclass{llncs}
\usepackage{graphicx}

\newcommand{\comment}[1]{}

\begin{document}
%
\title{An IDE to Build and Check Task Flow Models}


\author{Carlos Alberto Fernandez-y-Fernandez\inst{1} \and Jose Angel Quintanar Morales \inst{2} \and Hermenegildo Fernandez Santos \inst{2}}
\institute{Instituto de Computaci\'on, Universidad Tecnol\'ogica de la Mixteca, M\'exico 
\and Lab. de Inv. y Des. en Ing. de Soft., Universidad Tecnol\'ogica de la Mixteca, M\'exico  \\
\email{\{caff, joseangel, ps2010160001\}@mixteco.utm.mx}
}

\maketitle

\begin{abstract}

This paper presents the Eclipse plug-ins for the Task Flow model in
the Discovery Method. These plug-ins provide an IDE for the Task Algebra compiler and the model-checking tools.
The Task Algebra is the formal representation for the Task Model and it is based on simple and compound
tasks.  The model-checking techniques were developed to validate Task Models represented in the algebra. 

\end{abstract}
\begin{keywords} lightweight formal specification; software modelling; model-checking.

\end{keywords}

\section{Introduction}

There has been a steady take up in the use of formal calculi for software
construction over the last 25 years \cite{Bogdanov:03}, but mainly
in academia. Although there are some accounts of their use in industry
(basically in critical systems), the majority of software houses in
the \textquotedblleft{}real world\textquotedblright{} have preferred
to use visual modelling as a kind of \textquotedblleft{}semi-formal\textquotedblright{}
representation of software. A method is considered formal if it has
well-defined mathematical basis. Formal methods provide a syntactic
domain (i.e., the notation or set of symbols for use in the method),
a semantic domain (like its universe of objects), and a set of precise
rules defining how an object can satisfy a specification \cite{Wing:90}.
In addition, a specification is a set of sentences built using the
notation of the syntactic domain and it represents a subset of the
semantic domain. Spivey says that formal methods are based on mathematical
notations and \textquotedblleft{}they describe what the system must
do without saying how it is to be done\textquotedblright{} \cite{Spivey:89},
which applies to the non-constructive approach only. Mathematical
notations commonly have three characteristics: 
\begin{itemize}
\item conciseness - they represent complex facts of a system in a brief
space; 
\item precision - they can specify exactly everything that is intended; 
\item unambiguity - they do not admit multiple or conflicting interpretations. 
\end{itemize}
Essentially, a formal method can be applied to support the development
of software and hardware. This paper shows an IDE for modelling and checking task flow models using
a particular process algebra, called Task Algebra, to characterise
the Task Flow models in the Discovery Method. The advantage is that
this will allow software engineers to use diagram-based design methods
that have a secure formal underpinning.

\subsection{The Discovery Method}

The Discovery Method is an object-oriented methodology proposed formally
in 1998 by Simons \cite{Simons:98,Simons:98b}; it is considered by
the author to be a method focused mostly on the technical process. The Discovery Method is organised into four phases;
Business Modelling, Object Modelling, System Modelling, and Software
Modelling (Simons, pers. comm.).The Business Modelling phase is task-oriented.
A task is defined in the Discovery Method as something that \textquotedblleft{}has
the specific sense of an activity carried out by stakeholders that
has a business purpose\textquotedblright{} (Simons, pers. comm.). This
task-based exploration will lead eventually towards the two kinds
of Task Diagrams: The Task Structure and Task Flow Diagrams. The workflow
is represented in the Discovery Method using the Task Flow Diagram.
It depicts the order in which the tasks are realised in the business,
expressing also the logical dependency between tasks. While the notation
used in the Discovery Method is largely based on the Activity Diagram
of UML, it maintains consistently the labelled ellipse notations for
tasks. 

\subsection{The Task Flow models}

Even though Task Flow models could be represented using one of the
process algebras described above, a particular algebra was defined
with the aim of having a clearer translation between the graphical
model and the algebra. One of the main difficulties with applying
an existing process algebra was the notion that processes consist
of atomic steps, which can be interleaved. This is not the case in
the Task Algebra, where even simple tasks have a non-atomic duration
and are therefore treated as intervals, rather than atomic events.
A simple task in the Discovery Method \cite{Simons:98}
is the smallest unit of work with a business goal. A simple task is
the minimal representation of a task in the model. A compound task
can be formed by either simple or compound tasks in combination with
operators defining the structure of the Task Flow Model. In addition
to simple tasks and compound tasks, the abstract syntax also requires
the definition of three instantaneous events. These may form part
of a compound task in the abstract syntax.

\section{The Task Flow metamodel}

\subsection{The Task Algebra for Task Flow models}

The basic elements of the abstract syntax are: the simple task, which
is defined using a unique name to distinguish it from others; $\varepsilon$
representing the empty activity; and the success $\sigma$ and failure
$\varphi$ symbols, representing a finished activity.  Simple and compound tasks are combined using the operators that build up
the structures allowed in the Task Flow Model. The basic syntax structures
for the Task Flow Model are sequential composition, selection, parallel
composition, repetition, and encapsulation. \comment{ :
\begin{itemize}
\item \textbf{Sequential composition} defines the chronological order of
execution for a task or a group of tasks from the left to the right
and '$;$' is used as the operator. 
\item \textbf{Selection} is represented with the symbol '$+$' and it means
that there is a choice between the operands. 
\item \textbf{Parallel composition} defines the simultaneous execution of
the elements in the expression. It is represented by the symbol '$\parallel$. 
\item \textbf{Repetition} allows the reiteration of an expression in the
form of an until-loop and while-loop structure. It is represented
using the '$\mu x$' fixpoint. 
\item Finally, \textbf{encapsulation} is used to group a set of tasks and
structures. This constructs a compound task and is represented using
curly brackets '$\lbrace$' '$\rbrace$' . 
\end{itemize}
} The algebra definition is shown in table
\ref{abstract-syntax-table}. %
\begin{table}[h]
 \centering \begin{tabular}{lll}
$Activity::=$  & $\varepsilon$  & -- empty activity \tabularnewline
 & $|\sigma$  & -- succeed\tabularnewline
 & $|\varphi$  & -- fail \tabularnewline
 & $|Task$  & -- a single task \tabularnewline
 & $|Activity;Activity$  & -- a sequence of activity \tabularnewline
 & $|Activity+Activity$  & -- a selection of activity \tabularnewline
 & $|Activity\parallel Activity$  & -- parallel activity \tabularnewline
 & $|\mu x.(Activity;\varepsilon+x)$  & -- until-loop activity \tabularnewline
 & $|\mu x.(\varepsilon+Activity;x)$  & -- while-loop activity \tabularnewline
 &  & \tabularnewline
$Task::=$  & $Simple$  & -- a simple task \tabularnewline
 & $|{Activity}$  & -- encapsulated activity \tabularnewline
\end{tabular}\caption{abstract syntax definition}

\label{abstract-syntax-table} 
\end{table}

A task can be either a simple or a compound task. Compound tasks are
defined between brackets '$\lbrace$' and '$\rbrace$', and this is
also called encapsulation because it introduces a different context
for the execution of the structure inside it. Curly brackets are used
in the syntax context to represent diagrams and sub-diagrams but also
have implications for the semantics. Also, parentheses can be used to help comprehension or to change the
associativity of the expressions. Expressions associate to the right
by default.  More details of
the axioms can be seen in \cite{Fernandez:10}. 

\subsection{Model-checking}

A set of traces is the trace semantic representation for a Task Flow
Diagram. The verification of the diagram may be made in different
ways. The simplest operations could be performed by set operators
but more operations may be applied over the traces using temporal
logic. Temporal logic has being extensively applied with specification
and verification of software. The set of traces, obtained from a task
algebra expression, may be used to verify some temporal and logical
properties within the specification expressed by the diagrams. For
this reason, a simple implementation of LTL was built. This LTL implementation
works over the trace semantics generated from a Task Algebra expression.
Because the trace semantics represent every possible path of the Task
Flow diagram expressed in the Task Algebra, it is straightforward
to use LTL formulas to quantify universally over all those paths.
In this section, some examples using Linear Temporal Logic (LTL) are
presented, to illustrate the reasoning capabilities of the LTL module.
LTL is a temporal logic, formed adding temporal operators to the predicate
calculus. These operators that can be used to refer to future states
with no quantification over paths. In addition, a CTL application
was built to test CTL theorems against expressions in the task algebra.
In this case, the application has to transform the traces in a tree
representation before applying the expression. While LTL formulas
express temporal properties over all undifferentiated paths, Computational
Tree Logic (CTL) also considers quantification over sets of paths.
CTL is a branching-time logic \cite{Clarke:86} and theorems in this
logic may also be tested against a set of traces obtained from a task
algebra expression, in the same way that LTL theorems were tested
above.

\section{A tool for formal specification of Task Flow models}
\subsection{Analysis of Integrated Development Environments (IDE)}

Through a search in surveys and articles published in
digital media, Eclipse is chosen as the top two
open source IDEs best positioned among developers. However, Eclipse showed a better performance due to the existence
of robust tools for the development of plug-in, as it has with the
Plug-in's Development Environment (PDE) which provides tools to create,
develop, test, debug, build and deploy Eclipse plug-ins, modules and
features to update the sites and products Riched Client Platform (RCP). PDE consists of three elements: 
\begin{itemize}
\item PDE User Interface (UI) for designing the user interface; 
\item PDE Tools Application Programming Interface (API Tooling) useful pieces of code to develop applications;
\item PDE Builder (Build), manager responsible for the administration of the plug-in. 
\end{itemize}

Besides all this, the GMF frameworks (Graphic Modeling Framework
- Framework for graphic editing) and Eclipse Modeling Framework (Eclipse
Modeling Framework, EMF), which facilitate the construction. We can get a highly functional visual editor using EMF to build a structured data model enriched by GMF editors. The main advantage is that  being all development based on building a structured model, the time spent on the maintenance phase will be substantially reduced.
\subsection{The architecture of the task model tool}
As mentioned above, our general architecture is based on the Eclipse framework.  The first component is able to model Task Flow diagrams and translate them into a metamodel formed by Task Algebra expressions. The resultant file containing the metamodel is used by the Task Algebra compiler in order to generate the trace semantics.

\begin{figure}[here]
\begin{centering}

\includegraphics[scale=0.65]{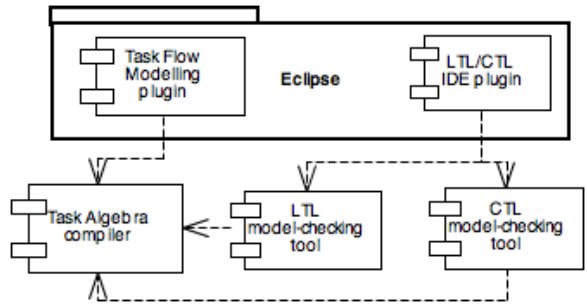}
\par
\end{centering}

\caption{Architecture of the Task Model Tool.}
\label{fig:architecture}
\end{figure}

In addition, the other component in Eclipse has the responsibility to receive LTL and CTL queries. The queries are sent to the relevant model-checking tool. Textual results are returned by the tool and have to be interpreted by LTL/CTL Eclipse plugin. Figure \ref{fig:architecture} shows the general dependency between the components of our project.

\section{Formal modelling made easy} 

\subsection{Design of the structured model}

\begin{figure}[here]
\begin{centering}
\includegraphics[scale=0.5]{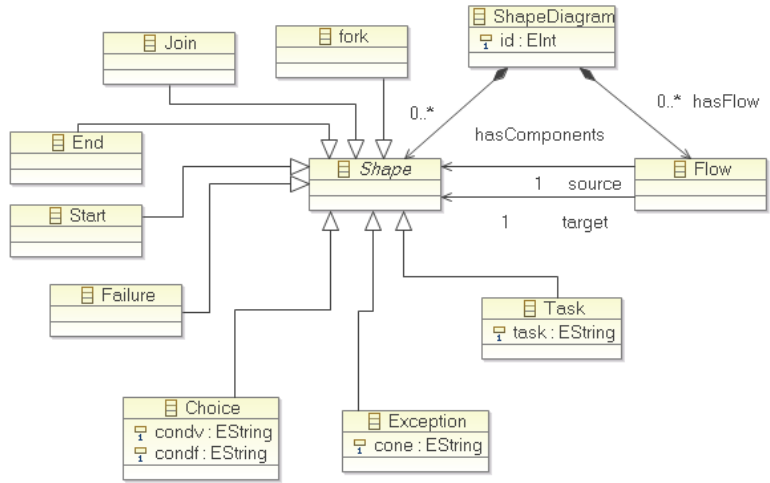}
\par\end{centering}
\caption{\label{fig:Modelo-de-clases}Class model for the Task model plug-in,based on GMF.}
\end{figure}

Once identified the use cases, classes were designed including the interaction
between different objects of the tool, we then proceeded to design the
structured model. This model is presented in Figure \ref{fig:Modelo-de-clases}. All development of the structured model is based on the use case
diagram, when we should be extra careful as it
migrates from an abstract model such as use cases
and results in a diagram from which one has the possibility of building
the computer application as such, in this case, set the application
logic. Note that only cover part of the user interaction.

\subsection{Development of the graphical model}

When the structured model is designed properly \cite{emf:2003,eclipse_dummies},
this can be transformed to the model graph. The model is a set of classes that represent real-world information.
In our case, the components which are integrated with diagrams. For
example, the Choice component, is associated with a specific behaviour, therefore
we need to store some additional information (i.e, this component implies information for the guards that
will trigger the flow).  All this without taking into account neither the
manner in which that information will be displayed nor the mechanisms
that make these data are part of the model; i.e., without regard to
any other entity within the plug-in.

\subsection{The domain model}

The domain model (or the model itself) is the set of classes resulting 
from analysing the components needed to design a task flow diagram.  Start, Task, Fork, Join, Exception, Failure,
Choice and End are the classes that were defined for the domain model. The domain model is not related to external information, we have an overview of the components of each one of its elements.

\subsection{The application model}

The application model is a set of classes that are related to the
domain model, are aware of the views and implement the necessary mechanisms
to notify the latter on the changes that might give the domain model.
The EMF framework, is responsible for this functionality, and which
interacts directly with the structured model; i.e., the model built on
EMF.

\subsection{The view domain}

The views are the set of classes that are responsible for showing
the user the information contained in the model. A view is associated
with a model. A view of the model gets only the information you need
to deploy and is updated each time the domain model changes through
notifications generated by the model of the application. GMF is responsible
for receiving such notifications and for generating visual feedback on
the plug-in.

\subsection{The driver}

The driver is an object that is responsible for directing the flow
of enforcement due to external messages and requests generations of the
algebra. From these messages, the controller modifies the model or
open and close views. The controller has access to the model and views,
but the view and the model are not aware of the existence of the controller.
The controller itself is the result of the implementation code from
the developer, which using GMF has the ability to interact with information
from the visual editor plug-in. This operation is given by
the $IWorkbenchWindowActionDelegate$ class implementation.

\subsection{Integration}

Finally when the two models have been integrated, we get almost all
of the user interface plug-in. It is at this point when we have to
develop the capabilities to manage graphics' performance and integration
with the components of the translator (i.e., the logic implementation,
where specific individual components). 

\begin{figure}[here]
\begin{centering}
\includegraphics[scale=0.35]{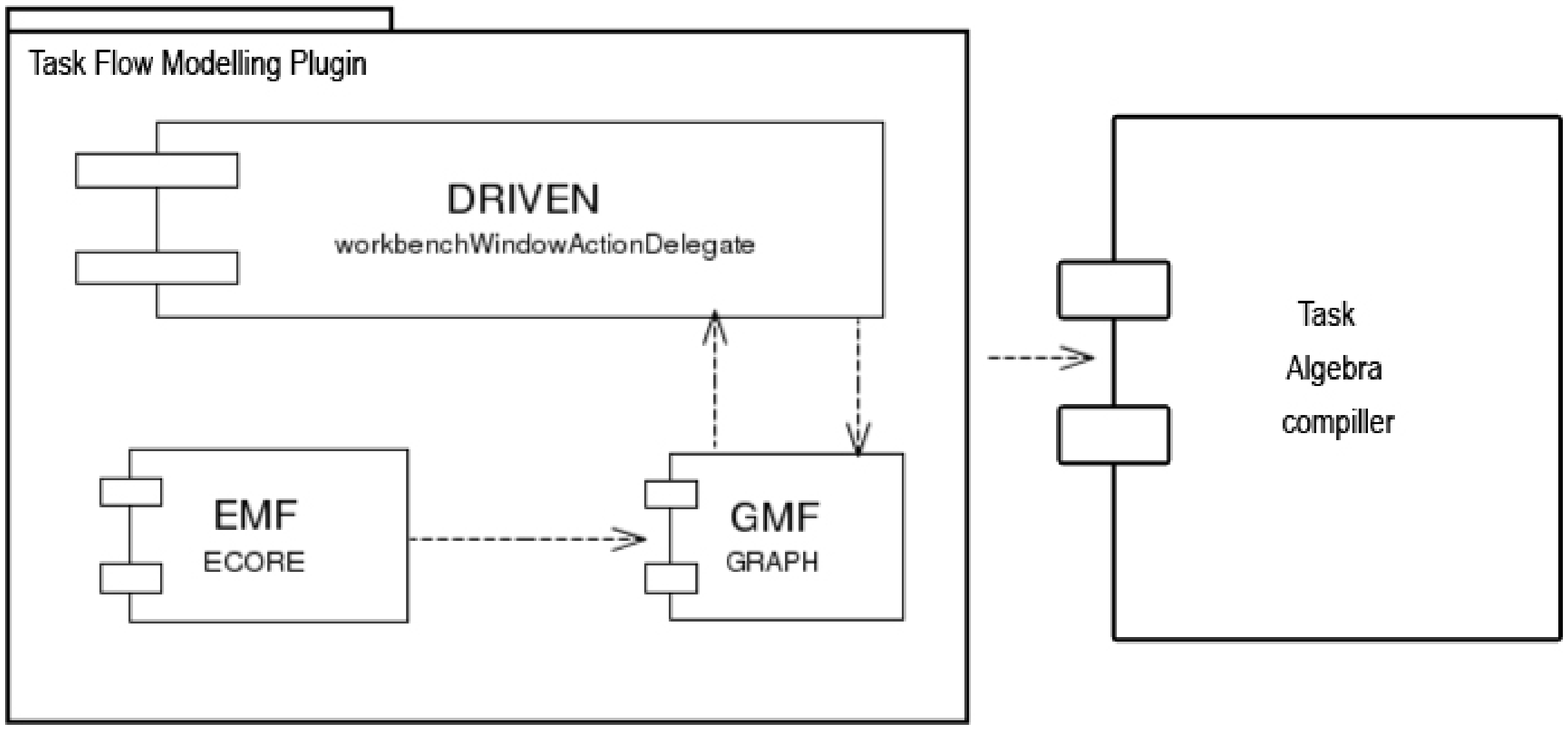}
\par\end{centering}
\caption{\label{fig:integracionl}View of integration and dependency of the plug-in for  development of tasks diagrams.}
\end{figure}

\subsection{Results}

\begin{figure}[here]
\begin{centering}
\includegraphics[scale=0.4]{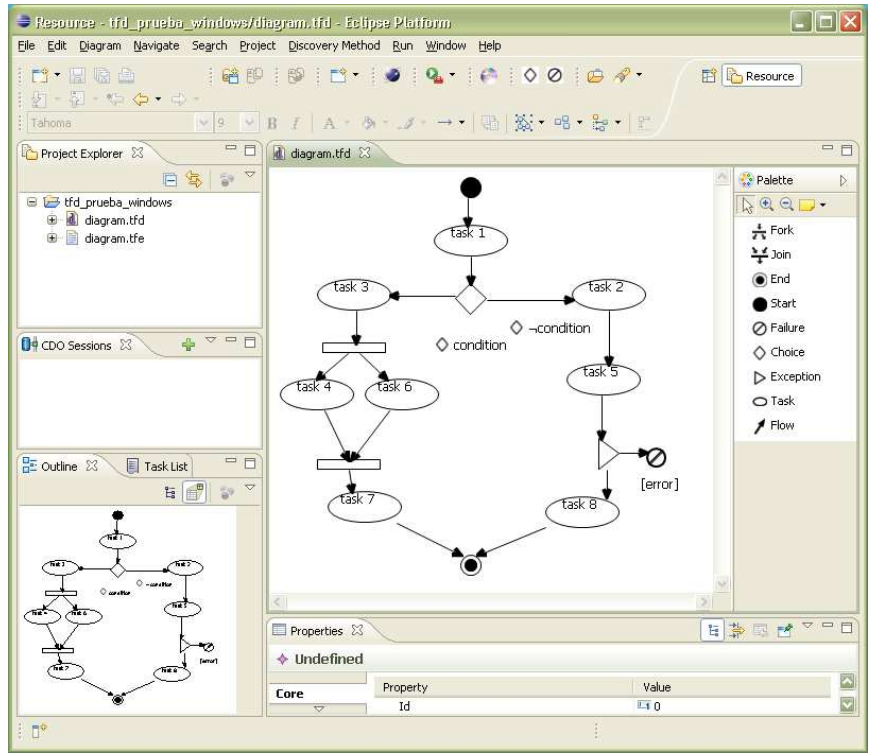}
\par\end{centering}
\caption{\label{fig:Vista-parcial-del}Partial view of user interface for the Task Model plug-in.}
\end{figure}

At this point we have obtained a comprehensive user interface, that is,
the party responsible for managing the design process diagrams. It
is worth noting that the code implementation has been rather small,
since everything is generated from structured model.  Up to this point
we have managed to cover about half of the project. Figure \ref{fig:Vista-parcial-del}
shows a screen user interface of this part of the project so far.

The development of application-based models implemented in the various
tools for creating plug-ins, as is the Plug-in Development Environment,
has resulted in optimization of time. The most important point is the possible modification, addition,
facilitation and exploration of the plug-in, because you can just
modify the structured model and its subsequent integration with GMF
model to make accurate changes, all without writing a single line
of code, so it is found that the design of the model implemented in
a tool is superior to developments made entirely in code.


\section{LTL and CTL model-checking IDE} 

\subsection{  Verification  Interface of Task Flow diagrams in the software specification}
Some factors influencing the development of quality software are: Understanding of requirements, proper modeling of the use cases, verification of models and development according to user needs. Task Flow diagrams from the Discovery Method are represented by a reduced and precise syntax.  The verification over the Task Flow diagrams is performed using temporal logic functions. The most common temporal logics are Linear Temporal Logic (LTL) and Computational Tree Logic (CTL)\cite{Chan-et-al:98}. 
The temporal logics are applied on an exhaustive set of states to see if a specification is true or not through time, it ensures verification of dynamic properties of a system without introducing time explicitly\cite{Gurfinkel-et-al:03}.   The Task Algebra proposed by Fernandez \cite{Fernandez:10} offers already the tools  (text mode) allowing you to verify Task Flow diagrams specified by the Discovery Method using temporal logic. This tool in text mode does not involve a visual representation of the operation and the logical transition of the model and it does not allow a full analysis of the results. The construction of an interface that allows to structure LTL/CTL queries and to graphically display results of the model verification represents the solution of the problem.

With the development of an interface to verify task diagrams, the user will have on hand a structured visual tool that lets him/her create logical expressions to refer to events in the algebraic model of work flow and display query results in a more meaningful and understandable way.  With the creation of these components the Task Algebra will become more accessible  and with the help of appropriate technologies it will represent a contribution to the specification and design phase in software development.

\subsection{Development Process}
The flow of activities in the design phase can be modeled by Task Flow diagrams, which in addition to its graphical representation has a formal syntactic model. The formal model of the task diagrams is the basis for verification of system properties.  The structure of a logical query(LTL/CTL) is complex, therefore it is necessary to assist it in the construction and comprehension of these expressions, as well as in the visualization of results.

\begin{figure}[here]
\centering
\includegraphics[scale=0.45]{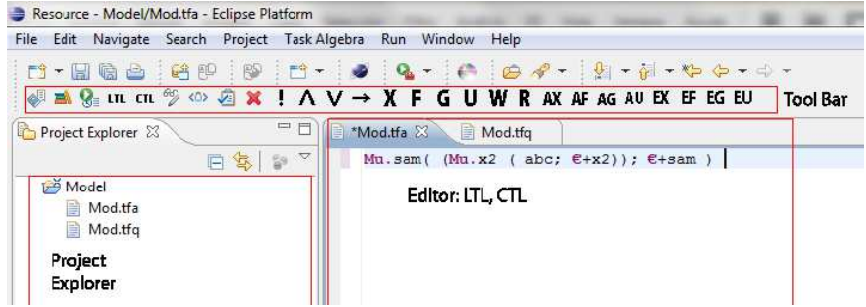}
\caption{\label{fig:Vista} Partial view of user interface for the model-checking plug-in: DSL Grammars, graphics elements and editor management.}
\end{figure}

Considering the ease of development, usage statistics and features offered in  development environments, the interface of verification will be integrated as a plug-in in the Eclipse development environment.  For best results, interface, testing and monitoring is necessary to take into consideration the following definitions for the task diagrams verification process:
 
\begin{itemize}
\item The plug-in should check the entry model that describes the task algebra.
\item There should be a check of logical expressions created (LTL and CTL syntax).
\item The test results should be displayed in an easy and simple way for user understanding.
\item The verification interface should be efficient and effective.

\end{itemize} 
Among the verification characteristics of the input model and the expressions syntax is used XText.
In order to verify the input model and the syntax of the expressions we use XText. With XText, domain-specific languages (DSL) can be created  in a formal and simple way. The framework supports the development of infrastructure in languages including compilers and interpreters and currently it has joined the Eclipse development environment. In interface development, Eclipse's core libraries such as org.eclipse.ui, org.eclipse.jface and org.eclipse.core are used. These packages allow to integrate icons and complete editor management, results in the interface development are shown in figure \ref{fig:Vista}. As we can see, the task diagrams verification interface consists of the following elements: module expressions, work area and input models.

\subsection{ Modules Interactivity and Results }
\begin{figure}[here]
\centering
\includegraphics[scale=0.38]{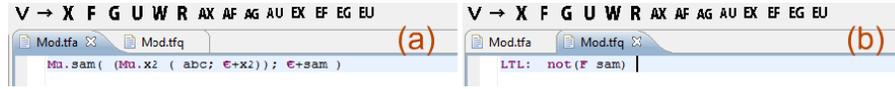}
\caption{\label{fig:Vista2} Partial view of the user interface for the model-checking plug-in: (a) describing a Task Flow diagram, (b) describing logical expressions. }
\end{figure}

The input for this plug-in is a Task Algebra expression representing the Task Flow metamodel (see Figure \ref{fig:Vista2}a).  This metamodel is used to generate the trace semantics needed to execute the query. 
A query construction is created and stored when the user builds LTL or CTL logical expressions (see Figure \ref {fig:Vista2}b).

The algebra model (tfa) and logical expressions created (tfq) are verified in continuous time using DSL grammars defined in the plugin (XText), which produces syntactically correct expressions. Combining the algebra model and the correct logical expressions, the verification of properties in the model is executed using the text mode tool described in \cite{Fernandez:10}. This part of the project is also responsible of the graphical display of the results. This is still a work in progress but it is considered relevant  in order to facilitate the interpretation of the query results. In particular, the CTL results are the most difficult to understand in their present form.

\section{Conclusions} 

Being Eclipse one of the most used environments for software development, we offer a tool that allows modelling and testing of software models that are defined usually in the specification phases.
Our research presented the Eclipse plug-ins for the Task Flow model in the Discovery Method. The task algebra is based on simple and compound tasks structured using operators such as sequence, selection, and
parallel composition. Recursion and encapsulation are also considered. The task algebra involves the definition of the denotational semantics for the task algebra, giving the semantics in terms of traces. Additionally, model-checking techniques were developed to validate Task Models represented in the algebra.
 
All of these was already available as console tools to prove the feasibility of the propose but, in order to be used by real-world developers, an IDE was necessary. With these tools, developers are not required to increase the quantity of artifacts when developing software. If developers create Task Flow diagrams, they will have an formal specification for their software which could improve communications using the unambiguous notation. In addition, using software model-checking in early stages may increase the confidence that goes from a correct definition to the final design. The plug-ins developed facilitate the formal specification of the Task Flow models   and the verification of these models in a visual and simple way.  The queries are structured visually and  with it the interpretation of results is even more simple. With this project the development time has been optimized and the quality of software has been guaranteed. In this project every module is easy to use and to understand for programmers due to its integration with Eclipse.

\section*{Acknowledgment}
This work has been funded by the UTM.

\bibliographystyle{splncs03} 
\bibliography{papertex}


\end{document}